\documentclass[twocolumn]{revtex4}
\usepackage{mathbbold}
\usepackage{amsfonts}
\usepackage{amsmath}
\usepackage{amssymb}
\usepackage{charter}
\usepackage{graphicx}

\begin{document}

\title{Tripartite entanglement and tripartite steering in three-qubit pure
states induced by vacuum--one-photon superpositions}
\author{Jian Wang, Huan Liu, Xue-feng Zhan and Xue-xiang Xu$^{\dag }$}
\affiliation{College of Physics and Communication Electronics, Jiangxi Normal University,
Nanchang 330022, China;\\
$^{\dag }$xuxuexiang@jxnu.edu.cn}

\begin{abstract}
Utilizing a tritter with variable parameter $T$ and induced by
vacuum--one-photon superpositions $\left\vert 0\right\rangle +\alpha
\left\vert 1\right\rangle $ with $\alpha =\left\vert \alpha \right\vert
e^{i\phi }$, we propose a scheme to prepare a class of three-qubit pure
states. These states take the form of $\left\vert \psi \right\rangle
_{123}=c_{0}\left\vert 000\right\rangle +c_{1}\left\vert 100\right\rangle
+c_{2}\left\vert 010\right\rangle +c_{3}\left\vert 001\right\rangle $. The
coefficients ($c_{0}$, $c_{1}$, $c_{2}$, and $c_{3}$) can be manipulated
through interaction parameters ($\left\vert \alpha \right\vert $, $\phi $,
and $T$). In line with Xie and Eberly's work[Phys. Rev. Lett. 127, 040403
(2021)], we investigate the genuine tripartite entanglement for $\left\vert
\psi \right\rangle _{123}$ by using the measure of concurrence fill. Drawing
on Hao \textit{et al.}'s research [Phys. Rev. Lett. 128, 120402 (2021)], we
examine tripartite steering for $\left\vert \psi \right\rangle _{123}$ under
certain measurements based on the uncertainty relations criterion. We
identify nine potential configurations exhibiting varying steerability
across different parameter spaces. It is important to highlight that, while
the state $\left\vert \psi \right\rangle _{123}$ exhibits entanglement,
steering remains unattainable in a substantial portion of the parameter
space.
\end{abstract}

\maketitle

\section{Introduction}

Entanglement is a key feature of quantum mechanics and plays an important
role in many quantum information protocols\cite{1,2,3}, including quantum
computation\cite{4}, quantum communication\cite{5} and quantum metrology\cite%
{6}. Previously, people paid more attention to studying bipartite
entanglement in two-party systems. To quantify the amount of entanglement,
they invented and developed a variety of entanglement measures, including
partial-norm\cite{7}, entanglement of formation\cite{8}, von Neumann entropy%
\cite{9}, normalized negativity\cite{10}, concurrence\cite{11}, and so on.
With the development of quantum technologies, more and more researchers
began to study multipartite entanglement (ME), existing in three-party or
even more-party systems. In general, ME can be divided as partial ME and
\textquotedblleft genuine\textquotedblright\ ME (GME). If a multipartite
state can be at least biseparable, then it is a partial ME but not a GME\cite%
{12}.

GME is crucial for quantum information and quantum technologies\cite{13}. In
general, a GME measure necessitates the following five requirements\cite%
{14,15}: (R1) It must assign the zero value to any product state or
biseparate state; (R2) It must assign a positive value to all nonbiseparate
states; (R3) It is convex; (R4) It is nonincreasing under local operations
and classical communication (LOCC); \ and (R5) It is invariant under local
unitary transformation. However, quantifying GME is still a challenge\cite%
{16} because most existing measures do not meet the \textquotedblleft
genuine\textquotedblright\ requirements. For example, some measures, such as
Schmidt measure by Eisert\ and Briege\cite{17}, or global entanglement by
Meyer\ and Wallach\cite{18}, will violate R1. While other measures, like the
three-tangle by Coffman \textit{et\ al.}\cite{19}, or a generalized form of
negativity by Jungnitsch \textit{et\ al.}\cite{20}, will violate (R2).

Recently, Xie and Eberly introduced a measure of genuine tripartite
entanglement (called \textquotedblleft concurrence fill\textquotedblright ),
which was defined as the square root of the area of concurrence triangle,
multiplying a constant factor\cite{21}. However, through a counterexample,
Ge \textit{et al.} pointed out that concurrence fill was a genuine
entanglement measure, but not an entanglement monotone\cite{22}. Afterwards,
they presented several faithful geometric measures for GME\cite{23}.

Einstein-Podolsky-Rosen (EPR) steering\cite{24,25}, which stipulates that
one observer can manipulate another party's state through local
measurements, is a crucial resource for various quantum applications\cite{26}%
. Typically, two methods are employed to explore multipartite steering: the
one-sided device-independent scenario\cite{27} and the steering correlation
between bipartitions\cite{28,29}. Key areas of studying multipartite EPR
steering include the monogamy\cite{30,31,32,33} and the shareability\cite{34}%
. Monogamy suggests that two observers cannot simultaneously steer the state
of a third party, while the shareability implies that two observers can
simultaneously steer a third observer. Over recent years, the monogamous
aspects of EPR steering have garnered significant attention in both
theoretical and experimental studies\cite{32}. To circumvent monogamous
relationships or eliminate monogamy constraints, researchers have uncovered
additional configurations of multipartite EPR steering by expanding the
number of measurement settings\cite{35}. Paul and Mukherjee recently
introduced explicit shareability relations using the violation of linear
steering inequality\cite{36}. Additionally, Hao \textit{et al.}
experimentally demonstrated various configurations of EPR steering
shareability using a three-qubit system\cite{37}.

Entanglement and steering, as resources, are pivotal in various quantum
protocols. A critical prerequisite is the distribution of these quantum
resources among multiple remote users within a network\cite{38,39}. Numerous
multi-qubit states, such as the two-qubit EPR state $\left( \left\vert
10\right\rangle +\left\vert 01\right\rangle \right) /\sqrt{2}$\cite{40},
three-qubit GHZ state $\left\vert GHZ\right\rangle =\left( \left\vert
000\right\rangle +\left\vert 111\right\rangle \right) /\sqrt{2}$\cite{41}
and three-qubit W state $\left\vert W\right\rangle =\left( \left\vert
100\right\rangle +\left\vert 010\right\rangle +\left\vert 001\right\rangle
\right) /\sqrt{3}$\cite{42}, have been extensively examined by assessing
their potential entanglement and steering\cite{43,44} for suitable
applications. In this study, we will introduce a class of three-qubit states
and analyze their tripartite entanglement and steering.

The paper is structured as follows. In Sec.II, we propose a scheme to
prepare a class of three-qubit pure states. In Sec.III, we explore "genuine"
tripartite entanglement by using the measure of concurrence fill. Sec.IV
delves into tripartite steering based on the uncertainty relations criterion
under specific measurement settings. Finally, Sec.V encapsulates the primary
findings.

\section{Preparation of three-qubit pure states}

In this section, we propose a scheme to prepare a class of three-qubit pure
states. As shown in Fig.1, we divide the entire process into the following
two stages.

\subsection{Stage 1: The preparation of the vacuum--one-photon superposition}

As proposed by Pegg, Phillips, and Barnett\cite{45}, the vacuum--one-photon
superposition (VOPS) $\left\vert 0\right\rangle +\alpha \left\vert
1\right\rangle $ can be prepared by utilizing a quantum scissor (QS) device.
The conceptual scheme is shown in Fig.1(a). The input coherent state $%
\left\vert \alpha \right\rangle $ is mixed on a balanced $BS_{2}$ with an
ancillary signal, and both outputs are measured using two\ single-photon
detectors. The ancillary signal is one of the two outputs of a single photon
passing another balanced $BS_{1}$, while the other signal is the output
VOPS. Of course, successful operation is heralded when a single photon is
detected at one detector and none at the\ other detector with perfect manner%
\cite{46}. It is important to note that this VOPS is truncated from the
input coherent state $\left\vert \alpha \right\rangle $ (by setting $\alpha
=\left\vert \alpha \right\vert e^{i\phi }$) and will subsequently be
normalized as $\left\vert \varepsilon \right\rangle =\omega _{0}\left\vert
0\right\rangle +\omega _{1}\left\vert 1\right\rangle $ with $\omega _{0}=1/%
\sqrt{1+\left\vert \alpha \right\vert ^{2}}$ and $\omega _{1}=\alpha /\sqrt{%
1+\left\vert \alpha \right\vert ^{2}}$. Very recently, Miranowicz \textit{et
al.} explored the nonclassicality of the VOPSs\cite{47}. In addition, the
prepared VOPS will serve as one of the input states of stage 2.

\subsection{Stage 2: Preparation of the three-qubit states under study}

As depicted in Fig.1(b), the kernel device is referred to as a tritter\cite%
{48,49} comprised of two consecutive BSs. We postulate the following: (i)
the initial BS is characterized by $\hat{B}_{12}(\pi /4)=e^{-\frac{\pi }{4}(%
\hat{a}_{1}^{\dag }\hat{a}_{2}-\hat{a}_{1}\hat{a}_{2}^{\dag })}$, satisfying
$\hat{B}_{12}\hat{a}_{1}^{\dag }\hat{B}_{12}^{\dag }=\frac{1}{\sqrt{2}}\hat{a%
}_{1}^{\dag }-\frac{1}{\sqrt{2}}\hat{a}_{2}^{\dag }$\ and $\hat{B}_{12}\hat{a%
}_{2}^{\dag }\hat{B}_{12}^{\dag }=\frac{1}{\sqrt{2}}\hat{a}_{1}^{\dag }+%
\frac{1}{\sqrt{2}}\hat{a}_{2}^{\dag }$, and (ii) the subsequent variable BS
is defined by$\ \hat{B}_{23}\left( \theta \right) =e^{-\theta (\hat{a}%
_{2}^{\dag }\hat{a}_{3}-\hat{a}_{2}\hat{a}_{3}^{\dag }}$), satisfying $\hat{B%
}_{23}\hat{a}_{2}^{\dag }\hat{B}_{23}^{\dag }=\sqrt{T}\hat{a}_{2}^{\dag }-%
\sqrt{1-T}\hat{a}_{3}^{\dag }$ and $\hat{B}_{23}\hat{a}_{3}^{\dag }\hat{B}%
_{23}^{\dag }=\sqrt{1-T}\hat{a}_{2}^{\dag }+\sqrt{T}\hat{a}_{3}^{\dag }$,
where $T=\cos ^{2}\theta $ $\in \lbrack 0,1]$. Notice that $\hat{a}%
_{j}^{\dag }$\ (and $\hat{a}_{j}$) denotes the creation (and annihilation)
operator of the $j$-th mode. Consequently, the tritter can be represented by$%
\ \hat{T}_{123}=\hat{B}_{23}\left( \theta \right) \hat{B}_{12}(\pi /4)$.
Therefore, we can generate a state yielding $\left\vert \psi \right\rangle
_{123}=\hat{T}_{123}\left\vert \varepsilon \right\rangle _{1}\left\vert
0\right\rangle _{2}\left\vert 0\right\rangle _{3}$ by injecting $\left\vert
\varepsilon \right\rangle $, $\left\vert 0\right\rangle $, $\left\vert
0\right\rangle $ into the corresponding input modes of the tritter. Upon
straightforward derivation, the prepared state can be explicitly articulated
as a three-qubit pure state%
\begin{equation}
\left\vert \psi \right\rangle _{123}=c_{0}\left\vert 000\right\rangle
+c_{1}\left\vert 100\right\rangle +c_{2}\left\vert 010\right\rangle
+c_{3}\left\vert 001\right\rangle ,  \label{1.1}
\end{equation}%
with $c_{0}=\omega _{0}$, $c_{1}=\omega _{1}/\sqrt{2}$, $c_{2}=-\omega _{1}%
\sqrt{T/2}$, and $c_{3}=\omega _{1}\sqrt{(1-T)/2}$. It is evident that the
state $\left\vert \psi \right\rangle _{123}$ correlates with three
interaction parameters (i.e., $\left\vert \alpha \right\vert $, $\phi $,\
and $T$). This state exhibits a hybrid form of both GHZ-class and W-class
states. Specifically, when $\alpha =0$, $\left\vert \psi \right\rangle
_{123} $ reduces to the simplest three-qubit product state $\left\vert
000\right\rangle $. However, if $\alpha \neq 0$\ and $T=0$, $\left\vert \psi
\right\rangle _{123}$\ transforms into a biseparable state $(\omega
_{0}\left\vert 00\right\rangle +\frac{\omega _{1}}{\sqrt{2}}\left\vert
10\right\rangle +\frac{\omega _{1}}{\sqrt{2}}\left\vert 01\right\rangle
)_{13}\otimes \left\vert 0\right\rangle _{2}$. Similarly, if $\alpha \neq 0$%
\ and $T=1$, $\left\vert \psi \right\rangle _{123}$\ becomes a biseparable
state $(\omega _{0}\left\vert 00\right\rangle +\frac{\omega _{1}}{\sqrt{2}}%
\left\vert 10\right\rangle +\frac{\omega _{1}}{\sqrt{2}}\left\vert
01\right\rangle )_{12}\otimes \left\vert 0\right\rangle _{3}$.
\begin{figure}[tbp]
\label{Fig1} \centering\includegraphics[width=0.95\columnwidth]{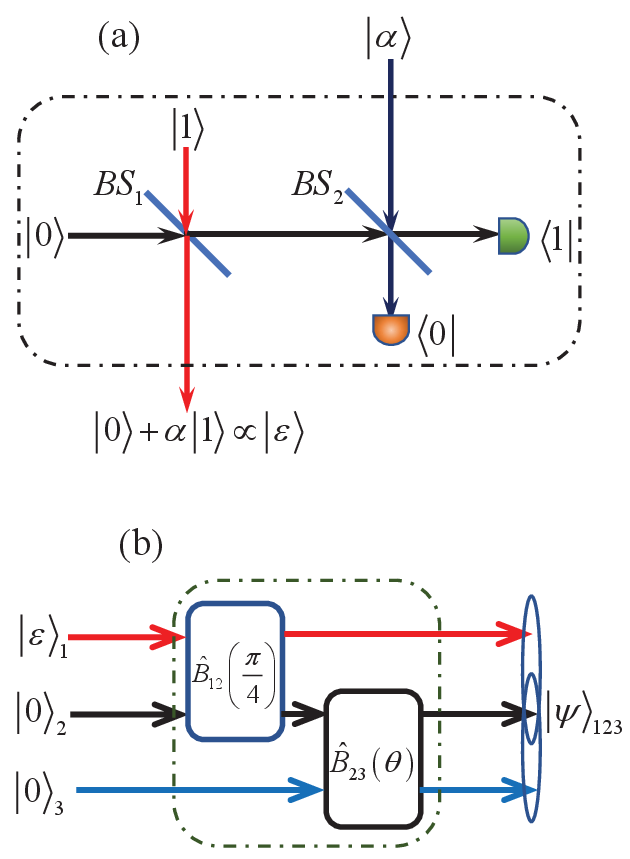}
\caption{(a) The preparation of the VOPS $\left\vert 0\right\rangle +\protect%
\alpha \left\vert 1\right\rangle \propto \left\vert \protect\varepsilon %
\right\rangle $ was achieved utilizing a QS device. This QS operation
consists of two BSs. (b) The preparation of the three-qubit state $%
\left\vert \protect\psi \right\rangle _{123}$ was executed using a tritter.}
\end{figure}

\section{Tripartite entanglement}

Concurrence is\ the most commonly used measure of entanglement. At the
beginning, this measure was mainly used to study the entanglement for
bipartite systems. In 2000, Coffman, Kundu, and Wootters first used the
concurrence to study the entanglement distribution for the pure three-qubit
states and showed the concurrence relation $C_{AB}^{2}+C_{AC}^{2}\leq
C_{A(BC)}^{2}$\cite{19}. Later, this measure was also generalized to study
ME, together with geometrical interpretation\cite{50}. In this paper, we
shall derive the $\mathcal{C}_{i(jk)}$-type concurrence (where $i,$ $j$ and $%
k$ are distinct values of $1$, $2$, or $3$) and investigate the
\textquotedblleft genuine\textquotedblright\ tripartite entanglement of $%
\left\vert \psi \right\rangle _{123}$. Herein, $\mathcal{C}_{i(jk)}$
represents the concurrence between a single party (inclusive of qubit $i$)
and another party (encompassing qubits $j$ and $k$).

\subsection{$\mathcal{C}_{i(jk)}$-type concurrence}

Utilizing the Schmidt decomposition, we can derive the Schmidt coefficients (%
$\sqrt{\lambda _{1}}$ and $\sqrt{\lambda _{2}}$) for any bipartite pure state%
\cite{51,52}. Consequently, the Schmidt weight can be ascertained through
\begin{equation}
Y=1-\sqrt{2(\lambda _{1}^{2}+\lambda _{2}^{2})-1}.  \label{2.1}
\end{equation}%
The concurrence can be computed using%
\begin{equation}
\mathcal{C}\left( Y\right) =\sqrt{Y\left( 2-Y\right) }.  \label{2.2}
\end{equation}

When $\left\vert \psi \right\rangle _{123}$ is treated as a bipartite state,
the corresponding Schmidt coefficients can be deduced (refer to Appendix A)
and the $\mathcal{C}_{i(jk)}$-type concurrence can be calculated according
to Eqs.(\ref{2.1}) and (\ref{2.2}). The primary findings are presented as
follows:

\textit{Case 1(23):} In the bipartite scenario involving qubit $1$ and pair $%
23$, we observe that:

\begin{equation}
Y_{1(23)}=1-\frac{\sqrt{\allowbreak 2\left\vert \alpha \right\vert ^{2}+1}}{%
\left\vert \alpha \right\vert ^{2}+1},  \label{2.3}
\end{equation}%
and%
\begin{equation}
\mathcal{C}_{1(23)}=\frac{\left\vert \alpha \right\vert ^{2}}{1+\left\vert
\alpha \right\vert ^{2}}\equiv \Omega .  \label{2.4}
\end{equation}

\textit{Case 2(31):} In the bipartite scenario involving qubit $2$ and pair $%
31$, we observe that:%
\begin{equation}
Y_{2(31)}=1-\frac{\sqrt{\allowbreak \left( 1-T\right) ^{2}\left\vert \alpha
\right\vert ^{4}+2\left\vert \alpha \right\vert ^{2}+1}}{\left\vert \alpha
\right\vert ^{2}+1},  \label{2.5}
\end{equation}%
$\allowbreak $and%
\begin{equation}
\mathcal{C}_{2(31)}=\Omega \sqrt{T\left( 2-T\right) }.  \label{2.6}
\end{equation}

\textit{Case 3(12):} In the bipartite scenario involving qubit $3$ and pair $%
12$, we observe that:%
\begin{equation}
Y_{3(12)}=1-\frac{\sqrt{T^{2}\left\vert \alpha \right\vert ^{4}+2\left\vert
\alpha \right\vert ^{2}+1}}{\left\vert \alpha \right\vert ^{2}+1},
\label{2.7}
\end{equation}%
and%
\begin{equation}
\mathcal{C}_{3(12)}=\Omega \sqrt{1-T^{2}}.  \label{2.8}
\end{equation}

Clearly, all $\mathcal{C}_{i(jk)}$s are contingent upon $\left\vert \alpha
\right\vert $\ and $T$, yet remain unaffected by $\phi $. Subsequently, we
will delve into the tripartite entanglement for $\left\vert \psi
\right\rangle _{123}$, utilizing the measure associated with $\mathcal{C}%
_{i(jk)}$-type concurrence.

\subsection{Concurrence triangles and fill}

In principle, any class of ME is linked to a geometric object, specifically
an entanglement polytope\cite{53}. Without a doubt, we can verify $\mathcal{C%
}_{i(jk)}\leq \mathcal{C}_{j(ki)}+\mathcal{C}_{k(ij)}$ for $\left\vert \psi
\right\rangle _{123}$ as per Qian, Alonso, and Eberly\cite{54}.\
Concurrently, we can also confirm%
\begin{equation}
\mathcal{C}_{i(jk)}^{2}\leq \mathcal{C}_{j(ki)}^{2}+\mathcal{C}_{k(ij)}^{2}.
\label{3.2}
\end{equation}%
for $\left\vert \psi \right\rangle _{123}$, following the method of Zhu and
Fei\cite{33}.

In accordance with Xie and Eberly\cite{21}, a concurrence triangle is
constructed by defining $s_{1}=\mathcal{C}_{1(23)}^{2}$, $s_{2}=\mathcal{C}%
_{2(31)}^{2}$, and $s_{3}=\mathcal{C}_{3(12)}^{2}$ as its three sides. It is
well established that the area of a triangle with side lengths ($s_{1}$, $%
s_{2}$, $s_{3}$) and perimeter $l=s_{1}+s_{2}+s_{3}$ can be computed using
Heron's formula $\mathcal{A}=\frac{1}{4}[l\left( l-2s_{1}\right) \left(
l-2s_{2}\right) \left( l-2s_{3}\right) ]^{1/2}$. For $\left\vert \psi
\right\rangle _{123}$, we can obtain the triangle area

\begin{equation}
\mathcal{A}_{\left\vert \psi \right\rangle _{123}}=\Omega ^{4}T\left(
1-T\right) \sqrt{1+T\left( 1-T\right) }.  \label{3.4}
\end{equation}%
and the concurrence fill\cite{21,22}%
\begin{equation}
F_{123}=F(\left\vert \psi \right\rangle _{123})=\sqrt{\frac{4}{\sqrt{3}}%
\mathcal{A}_{\left\vert \psi \right\rangle _{123}}}.  \label{3.5}
\end{equation}%
That is, the concurrence fill is just the square root of the triangle area
by multiplying $\sqrt{4/\sqrt{3}}$. Obviously, $\mathcal{A}_{\left\vert \psi
\right\rangle _{123}}$\ and $F_{123}$\ are dependent on $\left\vert \alpha
\right\vert $\ and $T$, but are independent of $\phi $.

In Fig.2, we give a table for the possible cases of $\left\vert \psi
\right\rangle _{123}$, accompanying their corresponding conditions,
concurrence triangles, areas, and fills. As highlighted by Dur \textit{et al.%
}\cite{12}, all three-qubit states can be categorized into three distinct
classes: product states, biseparable states, and nonbiseparable states. For $%
\left\vert \psi \right\rangle _{123}$, when $\left\vert \alpha \right\vert
=0 $, the triangle is simplified to a single dot due to $s_{1}=s_{2}=s_{3}=0$%
, resulting in an area $\mathcal{A}=0$ and $F_{123}=0$. When $\left\vert
\alpha \right\vert \neq 0$\ and $T=0$, the triangle is simplified to a line
due to $s_{1}=s_{3}=\Omega ^{2}>0$ and $s_{2}=0$, leading to an area $%
\mathcal{A}=0$ and $F_{123}=0$. When $\left\vert \alpha \right\vert \neq 0$\
and $T=1$, the triangle is simplified to a line due to $s_{1}=s_{2}=\Omega
^{2}>0$ and $s_{3}=0$, resulting in an area $\mathcal{A}=0$ and $F_{123}=0$.
Only when $\left\vert \alpha \right\vert \neq 0$ and $T\neq 0$ (or $1$),
does the triangle maintain its true form with appropriate $s_{1}$, $s_{2}$, $%
s_{3}>0$, accompanied by an area $\mathcal{A}>0$ and $F_{123}>0$.
\begin{figure*}[tbp]
\label{Fig2} \centering\includegraphics[width=1.8\columnwidth]{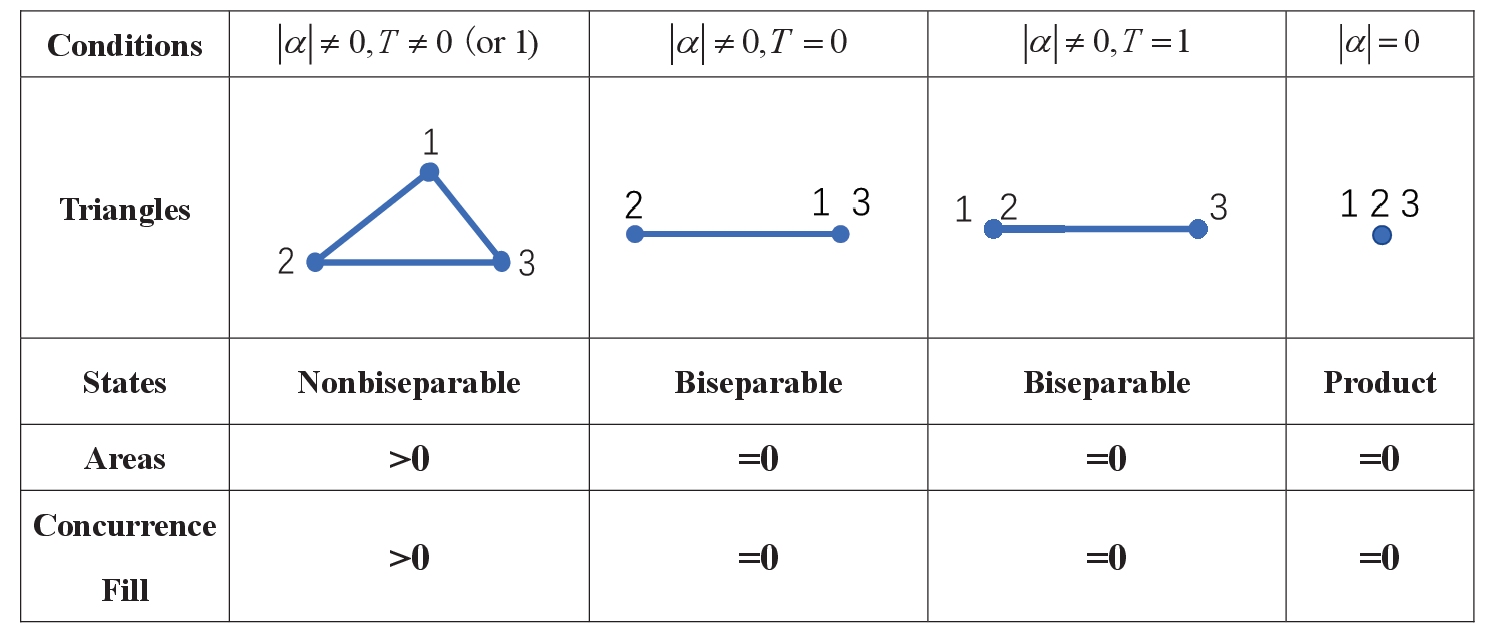}
\caption{The table of possible cases of $\left\vert \protect\psi %
\right\rangle _{123}$ and their corresponding conditions, concurrence
triangles, areas, and fills. }
\end{figure*}

As shown in Fig.3, the value of $\left\vert \alpha \right\vert $ (or $T$)
defines the shape (form) of the triangle for a fixed value of $T$ (or $%
\left\vert \alpha \right\vert $). In each sub-figure, one can see its
respective values for three side lengths, the area, and the concurrence
fill. Figures 3(a) to 3(c) depict the triangles with the same $\left\vert
\alpha \right\vert =5.5$ and different $T$ ($0.3$, $0.5$, and $0.7$),
accompanying different $F_{123}$ ($0.684394$, $0.752832$, and $0.684394$).
Figures 3(d) to 3(f) depict the triangles with same $T=0.5$ and different $%
\left\vert \alpha \right\vert $ ($1.5$, $2.5$, and $3.5$), accompanying
different $F_{123}$ ($0.3851$, $0.5971$, and $0.6867$). In Fig.4(a), we
present the contour plot of $F_{123}$\ in the ($\left\vert \alpha
\right\vert $, $T$) space. At the same time, we plot $F_{123}$\ as
functions\ of $\left\vert \alpha \right\vert $ for several different $T$ in
Fig.4(b) and $F_{123}$\ as functions of $T$\ for several different $%
\left\vert \alpha \right\vert $ in Fig.4(c). Obviously, $F_{123}$\ is a
symmetrical function of $T=0.5$ and reaches its maximal values at $T=0.5$
for each fixed $\left\vert \alpha \right\vert $. Meanwhile, $F_{123}$ is a
monotonically increasing function of $\left\vert \alpha \right\vert $\ for
each fixed $T$. At the limiting case of $\left\vert \psi \right\rangle
_{123} $ with $\left\vert \alpha \right\vert \rightarrow \infty $\ and $%
T=0.5 $, one can find a maximum value of $F_{123}$, i.e., $0.803428$. As
pointed out in Ref.\cite{21}, we also know $F_{123}\left( \left\vert
GHZ\right\rangle \right) =1$\ and $F_{123}\left( \left\vert W\right\rangle
\right) =8/9\doteq 0.889$. So, our considered state $\left\vert \psi
\right\rangle _{123}$ is still less entangled than the GHZ state and the W
state.

\begin{figure}[tbp]
\label{Fig3} \centering\includegraphics[width=0.8\columnwidth]{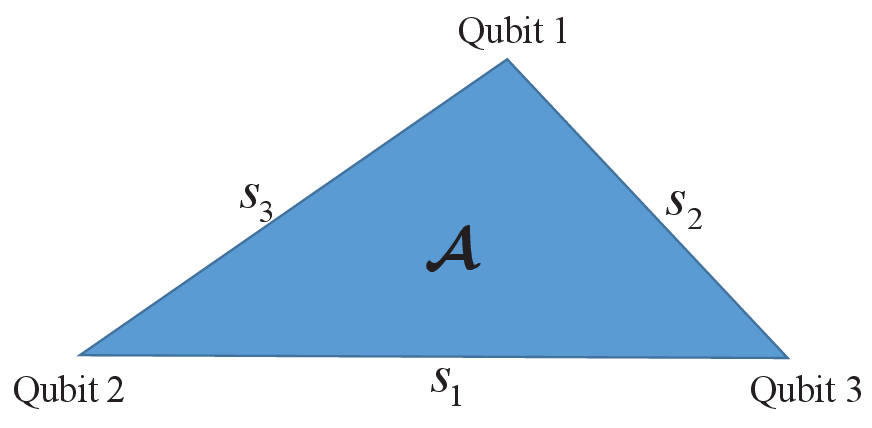} %
\centering\includegraphics[width=0.95\columnwidth]{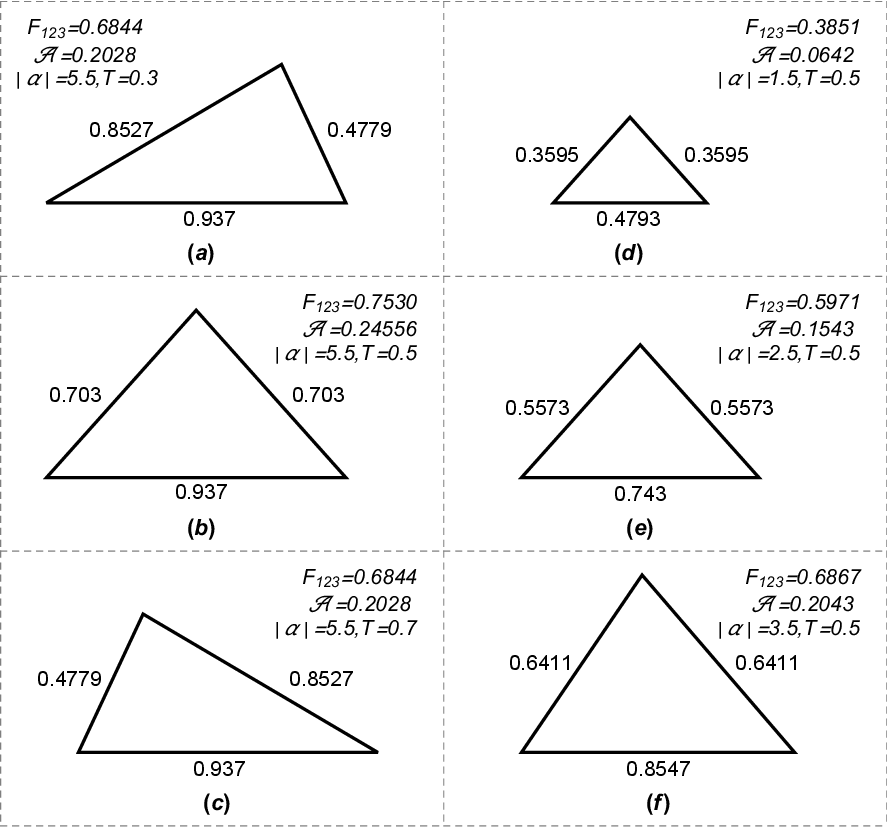}
\caption{Top: The concurrence triangle for a three-qubit state $\left\vert
\protect\psi \right\rangle _{123}$, with its three side lengths $s_{1}=%
\mathcal{C}_{1(23)}^{2}$, $s_{2}=\mathcal{C}_{2(31)}^{2}$ , and $s_{3}=%
\mathcal{C}_{3(12)}^{2}$. Bottom: Six triangles are depicted by taking $%
(\left\vert \protect\alpha \right\vert ,T)$ values with (a) $(5.5,0.3)$, (b)
$(5.5,0.5)$, (c) $(5.5,0.7)$, (d) $(1.5,0.5)$, (e) $(2.5,0.5)$, and (f) $%
(3.5,0.5)$, respectively. The corresponding areas and concurrence fills are
also provided.}
\end{figure}

\begin{figure*}[tbp]
\label{Fig43} \centering\includegraphics[width=2.0\columnwidth]{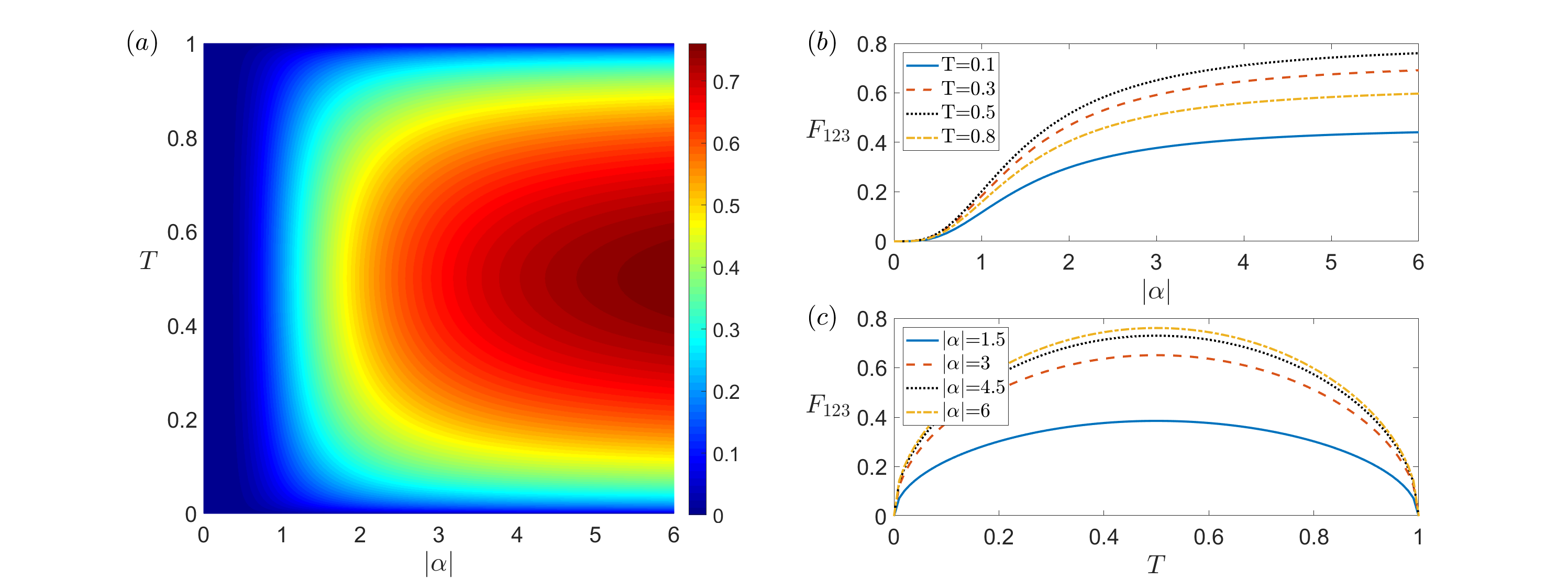}
\caption{(a) Contour plot of the concurrence fill $F_{123}\left( \left\vert
\protect\psi \right\rangle _{123}\right) $\ in the ($\left\vert \protect%
\alpha \right\vert $, $T$) space; (b) $F_{123}$ versus $\left\vert \protect%
\alpha \right\vert $\ with $T=0.1$, $0.3$, $0.5$, $0.8$; (c) $F_{123}$
versus $T$\ with $\left\vert \protect\alpha \right\vert =1.5$, $3$, $4.5$, $%
6 $.}
\end{figure*}

\subsection{Checking GME}

In the following, we shall check the five requirements for $F_{123}\left(
\left\vert \psi \right\rangle _{123}\right) $ one by one.

(R1) $F_{123}$ is zero when $\left\vert \psi \right\rangle _{123}$ is a
product state $\left\vert 000\right\rangle $\ for $\left\vert \alpha
\right\vert =0$ (see the fourth column in Fig.2), or when $\left\vert \psi
\right\rangle _{123}$ is a biseparate state\ for $\left\vert \alpha
\right\vert \neq 0$, $T=0$ (or $T=1$) (see the second and third column in
Fig.2).

(R2) $F_{123}$ is positive if $\left\vert \psi \right\rangle _{123}$ is a
nonbiseparate state\ for $\left\vert \alpha \right\vert \neq 0$, $T\neq 0$
(or $T\neq 1$) (see the first column in Fig.2).

(R3) As pointed out by Xie and Eberly\cite{21}, concurrence fill can be
constructed as the convex roof, i.e., $F_{123}\left( \rho \right)
=\min_{\{p_{i},\left\vert \psi _{i}\right\rangle
\}}\sum_{i}p_{i}F_{123}\left( \left\vert \psi _{i}\right\rangle \right) $,
over all possible decomposition $\rho =\sum_{i}p_{i}\left\vert \psi
_{i}\right\rangle \left\langle \psi _{i}\right\vert $. So, it is satisfying
the convex relation with $F_{123}(\sum_{i}p_{i}\left\vert \psi
_{i}\right\rangle \left\langle \psi _{i}\right\vert )\leq
\sum_{i}p_{i}F_{123}\left( \left\vert \psi _{i}\right\rangle \left\langle
\psi _{i}\right\vert \right) $. Obviously, this requirement is true for $%
\left\vert \psi \right\rangle _{123}$ by taking the equal sign.

(R4) Following the arguments in Refs.\cite{12,22,23} and through numerical
search in all parameter space, we find that $F_{123}\left( \left\vert \psi
\right\rangle _{123}\right) $ is nonincreasing under LOCC, i.e., $%
F_{123}\left( \Lambda _{LOCC}\left( \left\vert \psi \right\rangle
_{123}\right) \right) \leq F_{123}\left( \left\vert \psi \right\rangle
_{123}\right) $. The details on the LOCC-monotonicity are provided in
Appendix B.

(R5) Since $\mathcal{C}_{i(jk)}$ can be also obtained in another way through
$\sqrt{2[1-\mathrm{Tr}\left( \rho _{i}^{2}\right) }$ [$\rho _{i}=\mathrm{Tr}%
_{jk}\left( \rho _{123}\right) $], it is clear that $F_{123}\left(
\left\vert \psi \right\rangle _{123}\right) $ is invariant under local
unitary operations.

Then, we say, $F_{123}\left( \left\vert \psi \right\rangle _{123}\right) $
is a proper genuine tripartite entanglement measure.

\section{Tripartite steering}

Inspired by the work of Hao \textit{et al.}\cite{37}, we shall analyze the
tripartite EPR steering in $\left\vert \psi \right\rangle _{123}$, by using
the uncertainty relations\cite{55} under specific measurement settings.

\subsection{Theoretical proposal and relation}

We assume that qubits 1, 2 and 3 are controlled by Alice, Bob and Charlie,
respectively, with three measurement settings $\left\{ \sigma _{x},\sigma
_{y},\sigma _{z}\right\} $. In detail, we define these observables as $%
A_{1}=\sigma _{x}^{\left( 1\right) },A_{2}=\sigma _{y}^{\left( 1\right)
},A_{3}=\sigma _{z}^{\left( 1\right) }$; $B_{1}=\sigma _{x}^{\left( 2\right)
},B_{2}=\sigma _{y}^{\left( 2\right) },B_{3}=\sigma _{z}^{\left( 2\right) }$%
; $C_{1}=\sigma _{x}^{\left( 3\right) },C_{2}=\sigma _{y}^{\left( 3\right)
},C_{3}=\sigma _{z}^{\left( 3\right) }$, where%
\begin{equation}
\sigma _{x}^{\left( k\right) }=\left(
\begin{array}{cc}
0 & 1 \\
1 & 0%
\end{array}%
\right) ,\sigma _{y}^{\left( k\right) }=\left(
\begin{array}{cc}
0 & -i \\
i & 0%
\end{array}%
\right) ,\sigma _{z}^{\left( k\right) }=\left(
\begin{array}{cc}
1 & 0 \\
0 & -1%
\end{array}%
\right)  \label{4.1}
\end{equation}%
denote the standard Pauli spin operators for the $k$th qubit.

In the subsequent sections, we define the uncertainty of an observable $X$
on a state $\rho $ as the variance $\delta ^{2}X=\left\langle
X^{2}\right\rangle -\left\langle X\right\rangle ^{2}$. Here, $\left\langle
X\right\rangle $ represents the expectation value of $X$, calculated as $%
\left\langle X\right\rangle =\mathrm{Tr}\left( X\rho \right) $. Furthermore,
$C\left( X,Y\right) =\left\langle XY\right\rangle -\left\langle
X\right\rangle \left\langle Y\right\rangle $\ denotes the covariance between
observable $X$\ and observable $Y$. To ascertain the configuration of EPR
steering, we may employ the following criterion based on uncertainty
relations.

(1) \textit{Alice can steer Bob} if the inequality%
\begin{equation}
P_{AB}=\sum_{i}\delta ^{2}\left( \alpha _{i}^{(AB)}A_{i}+B_{i}\right) \geq
\min_{\rho _{B}}\sum_{i}\delta ^{2}B_{i}  \label{4.2}
\end{equation}%
is violated, where%
\begin{equation}
\alpha _{i}^{(AB)}=\left\{
\begin{array}{cc}
-\frac{C(A_{i},B_{i})}{\delta ^{2}A_{i}}, & \text{if }\delta ^{2}A_{i}\neq 0;
\\
-\frac{\delta ^{2}B_{i}}{2C(A_{i},B_{i})}, & \text{if }\delta
^{2}A_{i}=0,C(A_{i},B_{i})\neq 0; \\
0, & \text{if }\delta ^{2}A_{i}=0,C(A_{i},B_{i})=0.%
\end{array}%
\right.  \label{4.3}
\end{equation}

(2) \textit{Bob can steer Alice} if the inequality%
\begin{equation}
P_{BA}=\sum_{i}\delta ^{2}\left( \beta _{i}^{(BA)}B_{i}+A_{i}\right) \geq
\min_{\rho _{A}}\sum_{i}\delta ^{2}A_{i}  \label{4.4}
\end{equation}%
is violated, where%
\begin{equation}
\beta _{i}^{(BA)}=\left\{
\begin{array}{cc}
-\frac{C(A_{i},B_{i})}{\delta ^{2}B_{i}}, & \text{if }\delta ^{2}B_{i}\neq 0;
\\
-\frac{\delta ^{2}A_{i}}{2C(A_{i},B_{i})}, & \text{if }\delta
^{2}B_{i}=0,C(A_{i},B_{i})\neq 0; \\
0, & \text{if }\delta ^{2}B_{i}=0,C(A_{i},B_{i})=0.%
\end{array}%
\right.  \label{4.5}
\end{equation}

(3) \textit{Alice can steer Charlie} if the inequality%
\begin{equation}
P_{AC}=\sum_{i}\delta ^{2}\left( \alpha _{i}^{(AC)}A_{i}+C_{i}\right) \geq
\min_{\rho _{C}}\sum_{i}\delta ^{2}C_{i}  \label{4.6}
\end{equation}%
is violated, where%
\begin{equation*}
\alpha _{i}^{(AC)}=\left\{
\begin{array}{cc}
-\frac{C(A_{i},C_{i})}{\delta ^{2}A_{i}}, & \text{if }\delta ^{2}A_{i}\neq 0;
\\
-\frac{\delta ^{2}C_{i}}{2C(A_{i},C_{i})}, & \text{if }\delta
^{2}A_{i}=0,C(A_{i},C_{i})\neq 0; \\
0, & \text{if }\delta ^{2}A_{i}=0,C(A_{i},C_{i})=0.%
\end{array}%
\right.
\end{equation*}

(4) \textit{Charlie can steer Alice} if the inequality%
\begin{equation}
P_{CA}=\sum_{i}\delta ^{2}\left( \gamma _{i}^{(CA)}C_{i}+A_{i}\right) \geq
\min_{\rho _{A}}\sum_{i}\delta ^{2}A_{i}  \label{4.7}
\end{equation}%
is violated, where%
\begin{equation}
\gamma _{i}^{(CA)}=\left\{
\begin{array}{cc}
-\frac{C(A_{i},C_{i})}{\delta ^{2}C_{i}}, & \text{if }\delta ^{2}C_{i}\neq 0;
\\
-\frac{\delta ^{2}A_{i}}{2C(A_{i},C_{i})}, & \text{if }\delta
^{2}C_{i}=0,C(A_{i},C_{i})\neq 0; \\
0, & \text{if }\delta ^{2}C_{i}=0,C(A_{i},C_{i})=0.%
\end{array}%
\right.  \label{4.8}
\end{equation}

(5) \textit{Bob can steer Charlie} if the inequality%
\begin{equation}
P_{BC}=\sum_{i}\delta ^{2}\left( \beta _{i}^{(BC)}B_{i}+C_{i}\right) \geq
\min_{\rho _{C}}\sum_{i}\delta ^{2}C_{i}  \label{4.9}
\end{equation}%
is violated, where%
\begin{equation}
\beta _{i}^{(BC)}=\left\{
\begin{array}{cc}
-\frac{C(B_{i},C_{i})}{\delta ^{2}B_{i}}, & \text{if }\delta ^{2}B_{i}\neq 0;
\\
-\frac{\delta ^{2}C_{i}}{2C(B_{i},C_{i})}, & \text{if }\delta
^{2}B_{i}=0,C(B_{i},C_{i})\neq 0; \\
0, & \text{if }\delta ^{2}B_{i}=0,C(B_{i},C_{i})=0.%
\end{array}%
\right.  \label{4.10}
\end{equation}

(6) \textit{Charlie can steer Bob} if the inequality%
\begin{equation}
P_{CB}=\sum_{i}\delta ^{2}\left( \gamma _{i}^{(CB)}C_{i}+B_{i}\right) \geq
\min_{\rho _{B}}\sum_{i}\delta ^{2}B_{i}  \label{4.11}
\end{equation}%
is violated, where%
\begin{equation}
\gamma _{i}^{(CB)}=\left\{
\begin{array}{cc}
-\frac{C(B_{i},C_{i})}{\delta ^{2}C_{i}}, & \text{if }\delta ^{2}C_{i}\neq 0;
\\
-\frac{\delta ^{2}B_{i}}{2C(B_{i},C_{i})}, & \text{if }\delta
^{2}C_{i}=0,C(B_{i},C_{i})\neq 0; \\
0, & \text{if }\delta ^{2}C_{i}=0,C(B_{i},C_{i})=0.%
\end{array}%
\right.  \label{4.12}
\end{equation}

Some analytical results for calculating $P_{AB}$, $P_{BA}$, $P_{AC}$, $%
P_{CA} $, $P_{BC}$, and $P_{CB}$ are\ listed in Appendix C. For our used
setting, we can get $\min_{\rho _{A}}\sum_{i}\delta ^{2}A_{i}=2$, $%
\min_{\rho _{B}}\sum_{i}\delta ^{2}B_{i}=2$, and $\min_{\rho
_{C}}\sum_{i}\delta ^{2}C_{i}=2$. Physically, if $P_{ij}<2$, then we say
that party-$i$ can steer party-$j$. In particular, we have $%
P_{AB}=P_{BA}=P_{AC}=P_{CA}=P_{BC}=P_{CB}=2$ in the limiting case of $%
\left\vert \alpha \right\vert =0$, corresponding to product state $%
\left\vert 000\right\rangle $.

\subsection{Numerical simulation and analysis}

Using the\ above analytical expressions from Eqs.(\ref{4.2}) to (\ref{4.12}%
), we make numerical simulation for the tripartite steering of $\left\vert
\psi \right\rangle _{123}$.

In Fig.5, we plot the feasible regions satisfying $P_{AB}<2$, $P_{BA}<2$, $%
P_{AC}<2$, and $P_{CA}<2$, in ($\left\vert \alpha \right\vert $, $\phi $, $T$%
) parameter space by setting $0\leq T\leq 1$, $0\leq \left\vert \alpha
\right\vert \leq 6$, and $0\leq \phi \leq \pi $. Note that all $P_{ij}$s are
periodic functions of $\phi $ with period $\pi /2$. Moreover, it is
symmetric with respect to $\phi =\pi /4$ in the range $\phi \in \left( 0,\pi
/2\right) $. However, no matter what parameter ($\left\vert \alpha
\right\vert $, $\phi $, $T$) values we choose, it is impossible to satisfy $%
P_{BC}<2$\ and $P_{CB}<2$. That is to say, regions with $P_{BC}<2$ and $%
P_{CB}<2$ are empty,\ which means that there is no steering between $B$\ and
$C$. Undoubtedly, each sub-figure in Fig.5 only shows its respective one-way
steerability. The common region in Figs.5(a) and 5(b), satisfying $P_{AB}<2$%
\ and $P_{BA}<2$ simultaneously, will exhibit two-way steering between $A$
and $B$. Similarly, the common region in Fig.5(c) and 5(d), satisfying $%
P_{AC}<2$\ and $P_{CA}<2$ simultaneously, will exhibit two-way steering
between $A$ and $C$. Moreover, the regions with no steering are different
for these sub-figures.

In Fig.6, we depict three ($\left\vert \alpha \right\vert $, $T$) planes by
maintaining $\phi =0$, $0.1\pi $, $0.25\pi $ and illustrate nine distinct
configurations of steerability relations for $\left\vert \psi \right\rangle
_{123}$. Meanwhile, these configurations are detailed in Table I and further
elucidated in Fig.7. The implication of each configuration (here abbreviated
as Cf.) can be explained as follows.

Cf.\textquotedblleft a\textquotedblright\ signifies that Alice, Bob, and
Charlie are unable to steer each other (no steering).

Cf.\textquotedblleft b\textquotedblright\ denotes that only Alice and Bob
can steer each other (two-way steering).

Cf.\textquotedblleft c\textquotedblright\ indicates that solely Alice can
steer Bob (one-way steering). In this configuration, Bob cannot be steered
by Alice and Charlie simultaneously (a monogamy).

Cf.\textquotedblleft d\textquotedblright\ represents that: (1) Alice and Bob
can steer each other (two-way steering), and (2) Alice can steer Charlie
(one-way steering).

Cf.\textquotedblleft e\textquotedblright\ signifies that: (1) Alice and Bob
can steer each other (two-way steering), and (2) Alice and Charlie can steer
each other (two-way steering). In this configuration, Bob and Charlie can
simultaneously steer Alice (a shareability)

Cf.\textquotedblleft f\textquotedblright\ suggests that: (1) Alice can steer
Bob (one-way steering), and (2) Alice can steer Charlie (one-way steering).

Cf.\textquotedblleft g\textquotedblright\ implies that only Alice can steer
Charlie (one-way steering). In this configuration, Charlie cannot be steered
by Alice and Bob simultaneously (a monogamy).

Cf.\textquotedblleft h\textquotedblright\ indicates that only Alice and
Charlie can steer each other (two-way steering).

Cf.\textquotedblleft i\textquotedblright\ signifies that: (1) Alice and
Charlie can steer each other (two-way steering), and (2) Alice can steer Bob
(one-way steering).

Our configurations, further illustrated in Fig.7 and Table I, can reflect
their respective steering relations. For example, when $\phi =0.1\pi $, $%
\left\vert \alpha \right\vert =3.5$, and $T=0.5$, we have $P_{AB}=1.6719(4)$%
, $P_{BA}=1.8397(6)$, $P_{AC}=1.6719(4)$, $P_{CA}=1.8397(6)$, $%
P_{BC}=2.1978(1)$, and $P_{CB}=2.1978(1)$. This case corresponds to Fig.7(e).

As examples, we depict all $P_{ij}$s as functions of one parameter by fixing
other two parameters of $\left\vert \psi \right\rangle _{123}$ in Fig.8.
Through solving $P_{ij}=2$, we can obtain the intersection points in each
sub-figure and divide different ranges. For each range, we can identify its
corresponding configuration. Figure 8(a) presents all $P_{ij}$s\ versus $%
\left\vert \alpha \right\vert \in \lbrack 0,6]$\ for $\phi =0.1\pi $\ and $%
T=0.3$, where the ranges of $\left\vert \alpha \right\vert \in \left(
0,1.19751\right) $, $\left( 1.19751,1.28267\right) $, $\left(
1.28267,1.94563\right) $, and $\left( 1.94563,\infty \right) $\ correspond
to configurations of Figs.7(a), 7(g), 7(h), and 7(i). Fig.8(b) presents all $%
P_{ij}$s\ versus $\phi \in \lbrack 0,\pi ]$\ for $\left\vert \alpha
\right\vert =0.2$\ and $T=0.3$, where the ranges of $\phi \in \left(
0,0.175126\right) $, $\left( 0.175126,0.266456\right) $, $\left(
0.266456,0.306136\right) $, and $\left( 0.306136,\pi /4\right) $\ correspond
to configurations of Figs.7(i), 7(h), 7(g), and 7(a). Here, we only analyze
the variations in the range $\phi \in \lbrack 0,\pi /4]$, because all $P_{ij}
$s are the periodic functions with period $\pi /2$ and symmetrical in each
period. Fig.8(c) presents all $P_{ij}$s\ versus $T\in \lbrack 0,1]$\ for $%
\phi =0.1\pi $\ and $\left\vert \alpha \right\vert =0.2$, where the ranges
of $T\in \left( 0,0.210711\right) $, $\left( 0.210711,0.271447\right) $, $%
\left( 0.271447,0.728553\right) $, $\left( 0.728553,0.789289\right) $, and $%
\left( 0.789289,1\right) $\ correspond to configurations of Figs.7(h), 7(g),
7(a), 7(c), and 7(b).

\begin{figure}[tbp]
\label{FigS1} \centering\includegraphics[width=1.0\columnwidth]{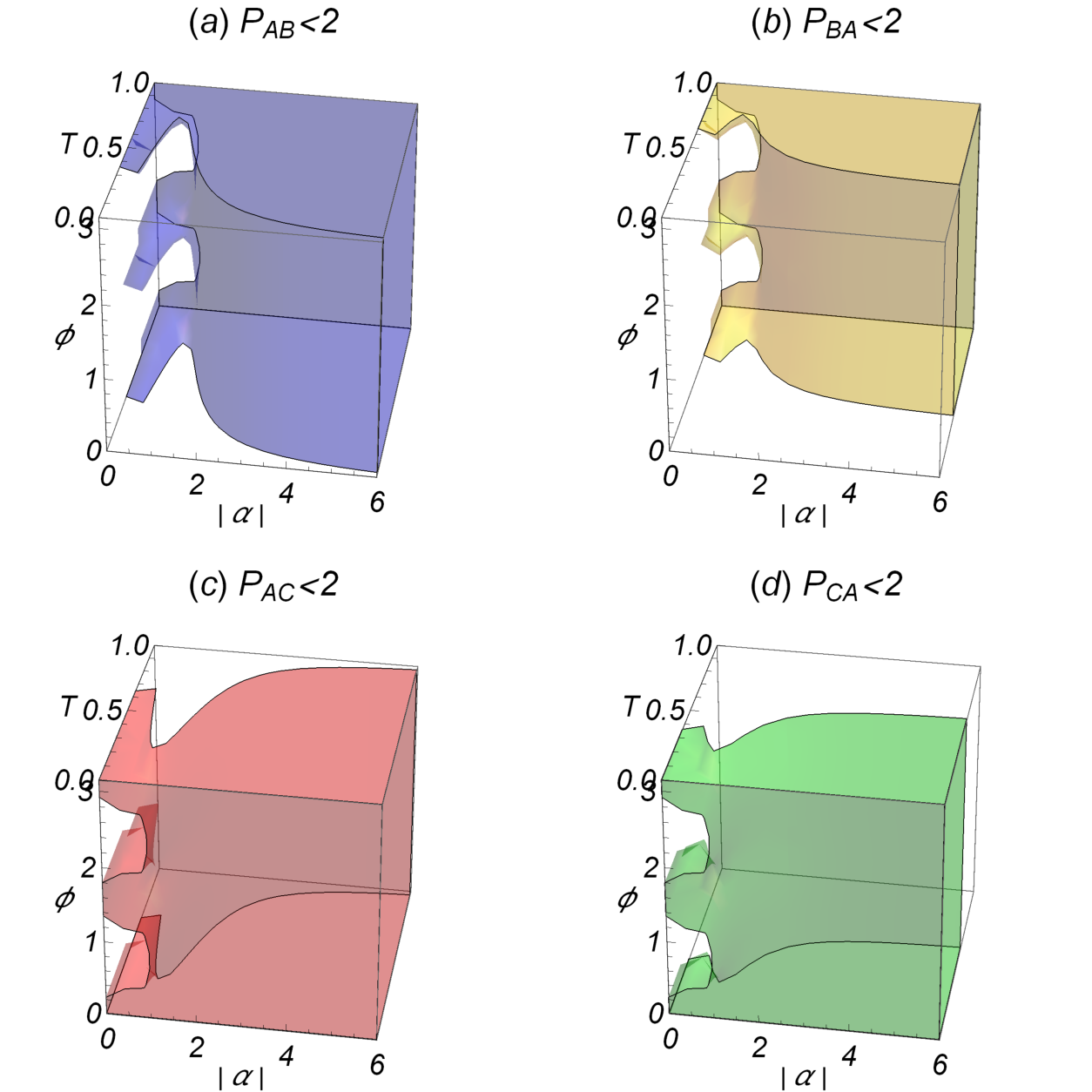}
\caption{The feasibility regions satisfying (a) $P_{AB}<2$, (b) $P_{BA}<2$,
(c) $P_{AC}<2$, (d) $P_{CA}<2$, in the parameters ($\left\vert \protect%
\alpha \right\vert $, $\protect\phi $, $T$) space with $\left\vert \protect%
\alpha \right\vert \in \lbrack 0,6]$, $\protect\phi \in \lbrack 0,\protect%
\pi ]$, $T\in \lbrack 0,1]$.}
\end{figure}

\begin{figure}[tbp]
\label{FigS2} \centering\includegraphics[width=1.0\columnwidth]{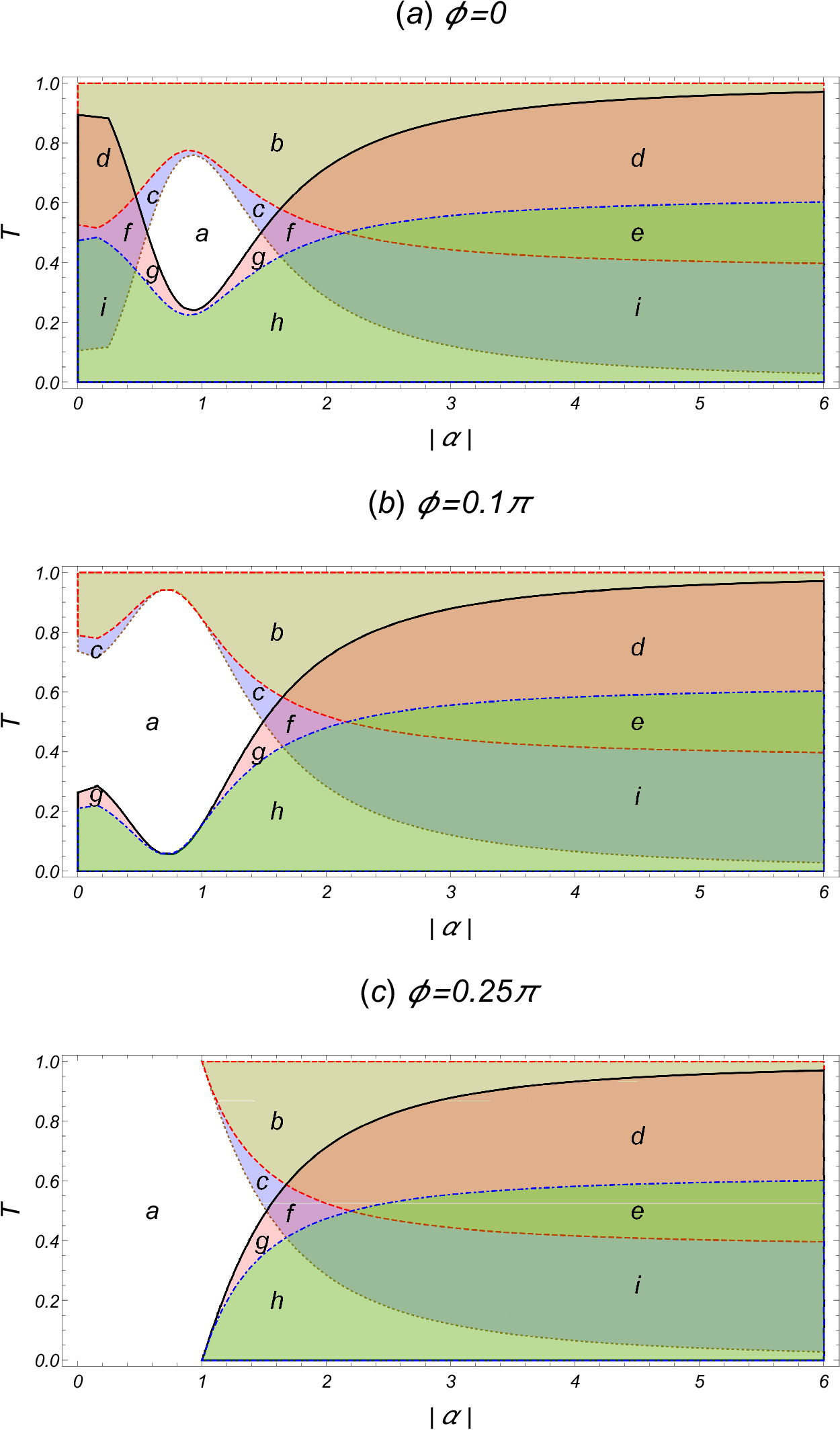}
\caption{Three ($\left\vert \protect\alpha \right\vert $, $T$) plains
illustrate nine distinct configurations of steerability for $\left\vert
\protect\psi \right\rangle _{123}$, characterized by (a) $\protect\phi =0$,
(b) $\protect\phi =0.1\protect\pi $, (c) $\protect\phi =0.25\protect\pi $,
respectively. The regions delineated by the colors correspond to these
varying configurations.}
\end{figure}

\begin{figure}[tbp]
\label{FigS3} \centering\includegraphics[width=1.0\columnwidth]{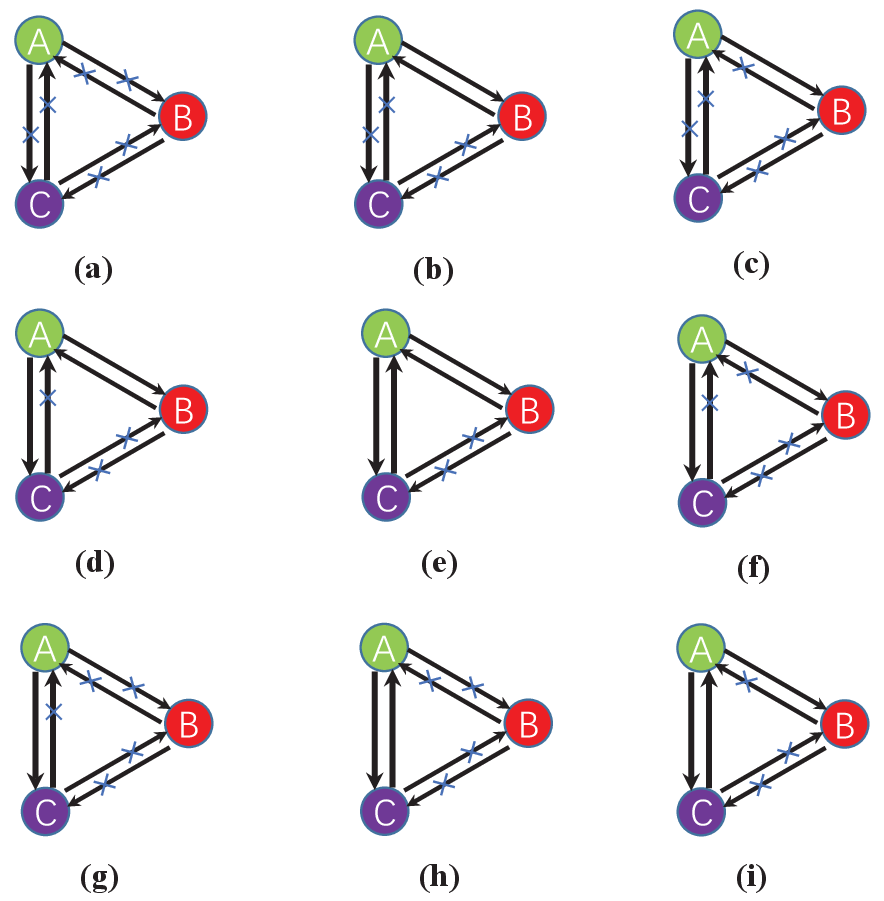}
\caption{The configurations of tripartite steerings, which are shared among
three observers (A, B, and C), correspond to regions delineated from (a) to
(i) in Fig.6.}
\end{figure}

\begin{figure}[tbp]
\label{FigS4} \centering\includegraphics[width=1.0\columnwidth]{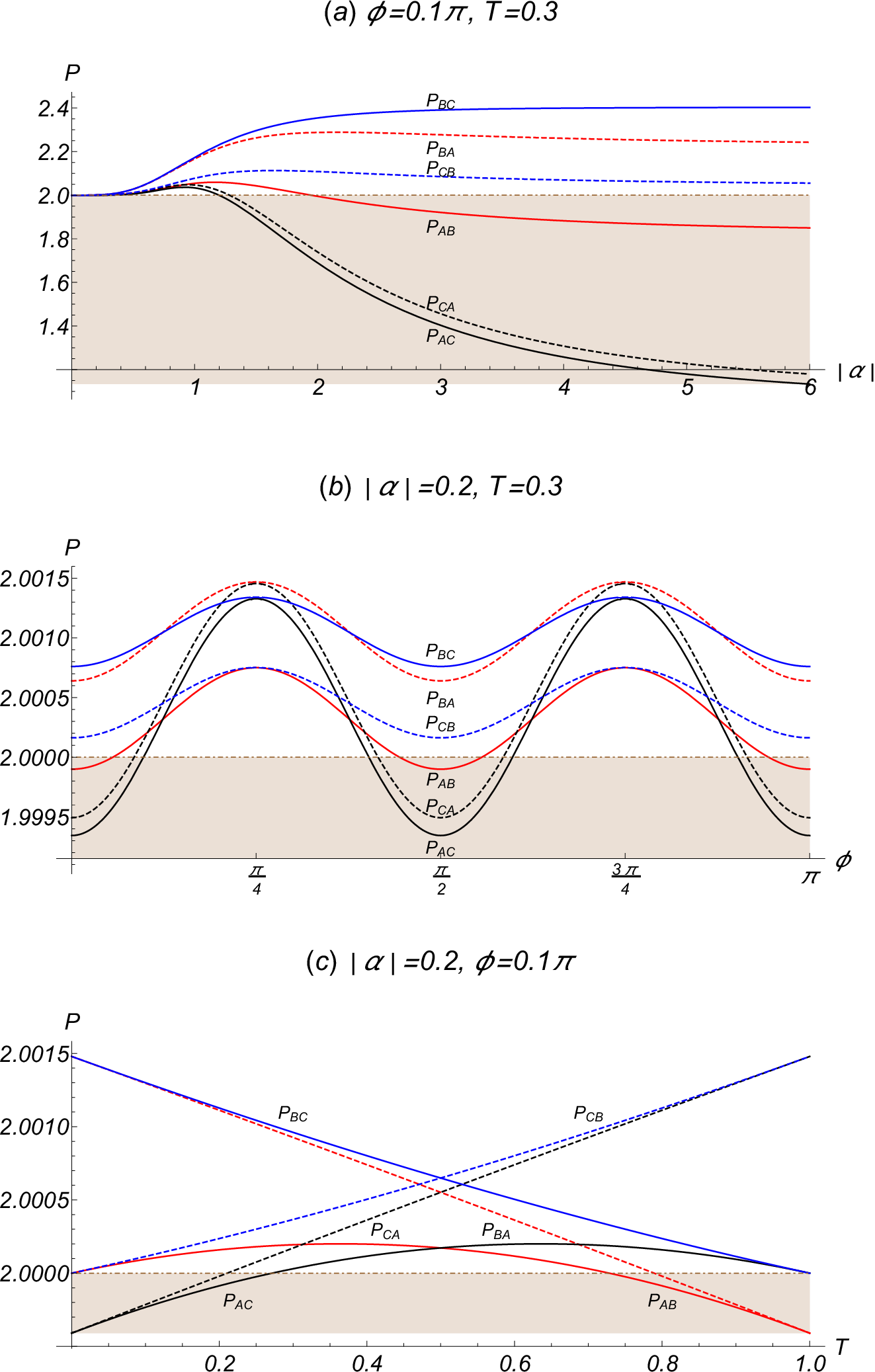}
\caption{ (a) $P$\ versus $\left\vert \protect\alpha \right\vert $ with
fixed $\protect\phi =0.1\protect\pi $ and $T=0.3$; (b) $P$\ versus $\protect%
\phi $ with fixed $\left\vert \protect\alpha \right\vert =0.2$ and $T=0.3$;
(c) $P$\ versus $T$ with fixed $\left\vert \protect\alpha \right\vert =0.2$
and $\protect\phi =0.1\protect\pi $.}
\end{figure}
\begin{table}[h]
\caption{Nine configurations (in Fig.6, Fig.7,\ and Table I) and their
respective steerings.}
\begin{center}
\begin{tabular}{|c||c|c|c|c|c|c|}
\hline\hline
configurations & $P_{AB}$ & $P_{BA}$ & $P_{AC}$ & $P_{CA}$ & $P_{BC}$ & $%
P_{CB}$ \\ \hline
a & $\geq 2$ & $\geq 2$ & $\geq 2$ & $\geq 2$ & $\geq 2$ & $\geq 2$ \\ \hline
b & $<2$ & $<2$ & $\geq 2$ & $\geq 2$ & $\geq 2$ & $\geq 2$ \\ \hline
c & $<2$ & $\geq 2$ & $\geq 2$ & $\geq 2$ & $\geq 2$ & $\geq 2$ \\ \hline
d & $<2$ & $<2$ & $<2$ & $\geq 2$ & $\geq 2$ & $\geq 2$ \\ \hline
e & $<2$ & $<2$ & $<2$ & $<2$ & $\geq 2$ & $\geq 2$ \\ \hline
f & $<2$ & $\geq 2$ & $<2$ & $\geq 2$ & $\geq 2$ & $\geq 2$ \\ \hline
g & $\geq 2$ & $\geq 2$ & $<2$ & $\geq 2$ & $\geq 2$ & $\geq 2$ \\ \hline
h & $\geq 2$ & $\geq 2$ & $<2$ & $<2$ & $\geq 2$ & $\geq 2$ \\ \hline
i & $<2$ & $\geq 2$ & $<2$ & $<2$ & $\geq 2$ & $\geq 2$ \\ \hline
\end{tabular}%
\end{center}
\end{table}

\section{Summary and conclusions}

In this study, we introduced a scheme for preparing a specific class of
three-qubit states and conducted a theoretical exploration of tripartite
entanglement and steering. These prepared states exhibit a hybrid form,
combining GHZ-like and W-like characteristics of three-qubit states. Through
the construction of a concurrence triangle, we demonstrated that our
three-qubit state possesses genuine tripartite entanglement. Utilizing
certain measurements and applying the uncertainty relations criterion, we
identified nine distinct configurations of tripartite steering. Notably,
most of these configurations adhere to shareability without being
constrained by monogamy. Throughout the paper, we provided comprehensive
analytical expressions and numerical results based on selected interaction
parameters. Upon thorough comparison and analysis, it was determined that
while the state exhibits entanglement, steering was unattainable in a
significant portion of the parameter space.

Our scheme has the advantages of using linear optics to prepare three-qubit
quantum states. With the help of BSs and photon-number-resolved detections,
we think, these states can be easily realized in experiments, especially by
virtue of the QS technique\cite{56,57,58}. Currently, there are many methods
to experimentally detect quantum correlations (including entanglement and
steering)\cite{59,60,61,62}. Moreover, some measurements (especially on
qubits) are relatively mature in the field of quantum information science%
\cite{63,64}. With current technology,\ we also think that detection for our
studied tripartite entanglement and steering can be implemented in
experiments.

\begin{acknowledgments}
This paper was supported by the National Natural Science Foundation of China
(Grants No. 12465004 and 11665013). We thank Bi-xuan Fan and Hong-chun Yuan
for the useful discussion.
\end{acknowledgments}

\textbf{Appendix A: Schmidt coefficients for calculating }$\mathcal{C}%
_{i(jk)}$\textbf{-type concurrence}

In accordance with the bipartite cases of $\left\vert \psi \right\rangle
_{123}$, we present the following Schmidt coefficients ($\lambda _{1}$s\ and
$\lambda _{2}$s) through the implementation of a Schmidt decomposition\cite%
{52}.

(1) \textit{Case 1(23):} If $\left\vert \psi \right\rangle _{123}$\ is
re-expressed as

\begin{equation}
\left\vert \psi \right\rangle _{123}=\left(
\begin{array}{cc}
\left\vert 0\right\rangle _{1} & \left\vert 1\right\rangle _{1}%
\end{array}%
\right) M_{1}\left(
\begin{array}{c}
\left\vert 00\right\rangle _{23} \\
\left\vert 10\right\rangle _{23} \\
\left\vert 01\right\rangle _{23} \\
\left\vert 11\right\rangle _{23}%
\end{array}%
\right)  \tag{A1}
\end{equation}%
with $M_{1}=\left(
\begin{array}{cccc}
c_{0} & c_{2} & c_{3} & 0 \\
c_{1} & 0 & 0 & 0%
\end{array}%
\right) $, then $\left\vert \psi \right\rangle _{123}$ can be further
decomposed into the Schmidt form like Eq.(17) in Ref.\cite{54} with the
Schmidt coefficients $\sqrt{\lambda _{1}^{\left( 1\right) }}$\ and $\sqrt{%
\lambda _{2}^{\left( 1\right) }}$, where%
\begin{align}
\lambda _{1}^{\left( 1\right) }& =\frac{1}{2}+\frac{\sqrt{2\left\vert \alpha
\right\vert ^{2}+1}}{2(\left\vert \alpha \right\vert ^{2}+1)},  \notag \\
\lambda _{2}^{\left( 1\right) }& =\frac{1}{2}-\frac{\sqrt{2\left\vert \alpha
\right\vert ^{2}+1}}{2(\left\vert \alpha \right\vert ^{2}+1)},  \tag{A2}
\end{align}%
are the eigenvalues of $M_{1}M_{1}^{\dag }$.$\allowbreak $

(2) \textit{Case 2(13):} If $\left\vert \psi \right\rangle _{123}$\ is
re-expressed as

\begin{equation}
\left\vert \psi \right\rangle _{123}=\left(
\begin{array}{cc}
\left\vert 0\right\rangle _{2} & \left\vert 1\right\rangle _{2}%
\end{array}%
\right) M_{2}\left(
\begin{array}{c}
\left\vert 00\right\rangle _{13} \\
\left\vert 10\right\rangle _{13} \\
\left\vert 01\right\rangle _{13} \\
\left\vert 11\right\rangle _{13}%
\end{array}%
\right)  \tag{A3}
\end{equation}%
with $M_{2}=\left(
\begin{array}{cccc}
c_{0} & c_{1} & c_{3} & 0 \\
c_{2} & 0 & 0 & 0%
\end{array}%
\right) $, then $\left\vert \psi \right\rangle _{123}$ can be further
decomposed into the Schmidt form like Eq.(17) in Ref.\cite{54} with the
Schmidt coefficients $\sqrt{\lambda _{1}^{\left( 2\right) }}$\ and $\sqrt{%
\lambda _{2}^{\left( 2\right) }}$, where%
\begin{align}
\lambda _{1}^{\left( 2\right) }& =\frac{1}{2}+\frac{\sqrt{\allowbreak
\allowbreak \left( 1-T\right) ^{2}\left\vert \alpha \right\vert
^{4}+2\left\vert \alpha \right\vert ^{2}+1}}{\allowbreak 2(\left\vert \alpha
\right\vert ^{2}+1)},  \notag \\
\lambda _{2}^{\left( 2\right) }& =\frac{1}{2}-\frac{\sqrt{\allowbreak
\allowbreak \left( 1-T\right) ^{2}\left\vert \alpha \right\vert
^{4}+2\left\vert \alpha \right\vert ^{2}+1}}{2(\left\vert \alpha \right\vert
^{2}+1)},  \tag{A4}
\end{align}%
are the eigenvalues of $M_{2}M_{2}^{\dag }$.

(3) \textit{Case 3(12):} If $\left\vert \psi \right\rangle _{123}$\ is
re-expressed as

\begin{equation}
\left\vert \psi \right\rangle _{123}=\left(
\begin{array}{cc}
\left\vert 0\right\rangle _{3} & \left\vert 1\right\rangle _{3}%
\end{array}%
\right) M_{3}\left(
\begin{array}{c}
\left\vert 00\right\rangle _{12} \\
\left\vert 10\right\rangle _{12} \\
\left\vert 01\right\rangle _{12} \\
\left\vert 11\right\rangle _{12}%
\end{array}%
\right)  \tag{A5}
\end{equation}%
with $M_{3}=\left(
\begin{array}{cccc}
c_{0} & c_{1} & c_{2} & 0 \\
c_{3} & 0 & 0 & 0%
\end{array}%
\right) $, then $\left\vert \psi \right\rangle _{123}$ can be further
decomposed into the Schmidt form like Eq.(17) in Ref.\cite{54} with the
Schmidt coefficients $\sqrt{\lambda _{1}^{\left( 3\right) }}$\ and $\sqrt{%
\lambda _{2}^{\left( 3\right) }}$, where%
\begin{align}
\lambda _{1}^{\left( 3\right) }& =\frac{1}{2}+\frac{\sqrt{T^{2}\left\vert
\alpha \right\vert ^{4}+2\left\vert \alpha \right\vert ^{2}+1}}{2(\left\vert
\alpha \right\vert ^{2}+1)},  \notag \\
\lambda _{2}^{\left( 3\right) }& =\frac{1}{2}-\frac{\sqrt{T^{2}\left\vert
\alpha \right\vert ^{4}+2\left\vert \alpha \right\vert ^{2}+1}}{2(\left\vert
\alpha \right\vert ^{2}+1)},  \tag{A6}
\end{align}%
are the eigenvalues of $M_{3}M_{3}^{\dag }$.

\textbf{Appendix B: Checking LOCC-monotonicity of concurrence fill}

Following the methods in Refs.\cite{12,22,23}, we check the
LOCC-monotonicity of $F_{123}\left( \left\vert \psi \right\rangle
_{123}\right) $.

First, we set $\ X_{1}=D_{1}V$\ and $X_{2}=D_{2}V$ as binary-outcome
positive-operator-valued measures (POVMs) satisfying $X_{1}^{\dag
}X_{1}+X_{2}^{\dag }X_{2}=\hat{I}$ ($\hat{I}$ is a $2\times 2$ identity
matrix), where%
\begin{equation}
D_{1}=\left(
\begin{array}{cc}
\sin \theta _{1} & 0 \\
0 & \sin \theta _{2}%
\end{array}%
\right) ,D_{2}=\left(
\begin{array}{cc}
\cos \theta _{1} & 0 \\
0 & \cos \theta _{2}%
\end{array}%
\right) ,  \tag{B1}
\end{equation}%
and%
\begin{equation}
V=\left(
\begin{array}{cc}
\cos \varkappa _{1} & -e^{i\varkappa _{2}}\sin \varkappa _{1} \\
\sin \varkappa _{1} & e^{i\varkappa _{2}}\cos \varkappa _{1}%
\end{array}%
\right)  \tag{B2}
\end{equation}%
with $\theta _{i}$, $\varkappa _{i}\in \lbrack -\pi ,\pi ]$.

Second, acting $X_{1}$\ and $X_{2}$\ on the mode-$1$ of $\left\vert \psi
\right\rangle _{123}$, we obtain%
\begin{equation}
\left\vert \psi ^{\left( 1\right) }\right\rangle _{123}=\frac{1}{\sqrt{p_{1}}%
}\left( X_{1}\otimes \hat{I}\otimes \hat{I}\right) \left\vert \psi
\right\rangle _{123},  \tag{B3}
\end{equation}%
and%
\begin{equation}
\left\vert \psi ^{\left( 2\right) }\right\rangle _{123}=\frac{1}{\sqrt{p_{2}}%
}\left( X_{2}\otimes \hat{I}\otimes \hat{I}\right) \left\vert \psi
\right\rangle _{123},  \tag{B4}
\end{equation}%
respectively. Here $p_{1}$\ and $p_{2}$\ denote their respective success
probability, satisfying $p_{1}+p_{2}=1$.

Third, we calculate $F(\left\vert \psi ^{\left( 1\right) }\right\rangle
_{123})$\ and $F(\left\vert \psi ^{\left( 2\right) }\right\rangle _{123})$
and check LOCC monotonicity. Through numerical search, we find that the
following inequality:%
\begin{equation}
F(\left\vert \psi \right\rangle _{123})-\sum_{i=1}^{2}p_{i}F(\left\vert \psi
^{\left( i\right) }\right\rangle _{123})\geq 0  \tag{B5}
\end{equation}%
is always satisfied in all possible parameters (including $\left\vert \alpha
\right\vert $, $\phi $, $T$, $\theta _{1}$, $\theta _{2}$, $\varkappa _{1}$,
and $\varkappa _{2}$) space. Since it is nonincreasing under the LOCC, we
say that the concurrence fill is entanglement monotone at least for our
considered state $\left\vert \psi \right\rangle _{123}$ and under our chosen
POVMs.

Moreover, similar conclusions can be obtained when the LOCC operations are
employed on mode-2 and mode-3.

$\allowbreak $\textbf{Appendix C: Analytical expressions of variances and
covariances}

Within the defined space spanned by \{$\left\vert 000\right\rangle $, $%
\left\vert 001\right\rangle $, $\left\vert 010\right\rangle $, $\left\vert
011\right\rangle $, $\left\vert 100\right\rangle $, $\left\vert
101\right\rangle $, $\left\vert 110\right\rangle $, $\left\vert
111\right\rangle $\}, the density operator $\rho _{123}=\left\vert \psi
\right\rangle _{123}\left\langle \psi \right\vert $ can be comprehensively
expanded into a matrix representation.%
\begin{widetext}
\begin{equation*}
\rho _{123}=\left(
\begin{array}{cccccccc}
\frac{1}{\left\vert \alpha \right\vert ^{2}+1} & \frac{\alpha \sqrt{2\left(
1-T\right) }}{2(\left\vert \alpha \right\vert ^{2}+1)} & -\frac{\alpha \sqrt{%
2T}}{2(\left\vert \alpha \right\vert ^{2}+1)} & 0 & \frac{\alpha \sqrt{2}}{%
2(\left\vert \alpha \right\vert ^{2}+1)} & 0 & 0 & 0 \\
\frac{\alpha ^{\ast }\sqrt{2\left( 1-T\right) }}{2(\left\vert \alpha
\right\vert ^{2}+1)} & \frac{\left\vert \alpha \right\vert ^{2}\left(
1-T\right) }{2(\left\vert \alpha \right\vert ^{2}+1)} & -\frac{\left\vert
\alpha \right\vert ^{2}\sqrt{T\left( 1-T\right) }}{2(\left\vert \alpha
\right\vert ^{2}+1)} & 0 & \frac{\left\vert \alpha \right\vert ^{2}\sqrt{1-T}%
}{2(\left\vert \alpha \right\vert ^{2}+1)} & 0 & 0 & 0 \\
-\frac{\alpha ^{\ast }\sqrt{2T}}{2(\left\vert \alpha \right\vert ^{2}+1)} & -%
\frac{\left\vert \alpha \right\vert ^{2}\sqrt{T\left( 1-T\right) }}{%
2(\left\vert \alpha \right\vert ^{2}+1)} & \frac{\left\vert \alpha
\right\vert ^{2}T}{2(\left\vert \alpha \right\vert ^{2}+1)} & 0 & -\frac{%
\left\vert \alpha \right\vert ^{2}\sqrt{T}}{2(\left\vert \alpha \right\vert
^{2}+1)} & 0 & 0 & 0 \\
0 & 0 & 0 & 0 & 0 & 0 & 0 & 0 \\
\frac{\alpha ^{\ast }\sqrt{2}}{2(\left\vert \alpha \right\vert ^{2}+1)} &
\frac{\left\vert \alpha \right\vert ^{2}\sqrt{1-T}}{2(\left\vert \alpha
\right\vert ^{2}+1)} & -\frac{\left\vert \alpha \right\vert ^{2}\sqrt{T}}{%
2(\left\vert \alpha \right\vert ^{2}+1)} & 0 & \frac{\left\vert \alpha
\right\vert ^{2}}{2(\left\vert \alpha \right\vert ^{2}+1)} & 0 & 0 & 0 \\
0 & 0 & 0 & 0 & 0 & 0 & 0 & 0 \\
0 & 0 & 0 & 0 & 0 & 0 & 0 & 0 \\
0 & 0 & 0 & 0 & 0 & 0 & 0 & 0%
\end{array}%
\right)
\end{equation*}
\end{widetext}

The following variances and covariances can be provided for the selected
observables.

(1) Three variances for Alice are

\begin{align}
\delta ^{2}A_{1}& =\frac{\left\vert \alpha \right\vert ^{4}+\left( 1-\cos
2\phi \right) \left\vert \alpha \right\vert ^{2}+1}{(\left\vert \alpha
\right\vert ^{2}+1)^{2}},  \notag \\
\delta ^{2}A_{2}& =\frac{\left\vert \alpha \right\vert ^{4}+\left( 1+\cos
2\phi \right) \left\vert \alpha \right\vert ^{2}+1}{(\left\vert \alpha
\right\vert ^{2}+1)^{2}},  \notag \\
\delta ^{2}A_{3}& =\frac{\left\vert \alpha \right\vert ^{4}+2\left\vert
\alpha \right\vert ^{2}}{(\left\vert \alpha \right\vert ^{2}+1)^{2}},
\tag{C1}
\end{align}%
which lead to%
\begin{equation}
\sum_{i}\delta ^{2}A_{i}=\frac{3\left\vert \alpha \right\vert
^{4}+4\left\vert \alpha \right\vert ^{2}+2}{(\left\vert \alpha \right\vert
^{2}+1)^{2}}.  \tag{C2}
\end{equation}

(2) Three variances for Bob are

\begin{align}
\delta ^{2}B_{1}& =\frac{\left\vert \alpha \right\vert ^{4}+\allowbreak
\left( 2-T-T\cos 2\phi \right) \left\vert \alpha \right\vert ^{2}+1}{%
(\left\vert \alpha \right\vert ^{2}+1)^{2}},  \notag \\
\delta ^{2}B_{2}& =\frac{\left\vert \alpha \right\vert ^{4}+\left( 2-T+T\cos
2\phi \right) \left\vert \alpha \right\vert ^{2}+1}{(\left\vert \alpha
\right\vert ^{2}+1)^{2}},  \notag \\
\delta ^{2}B_{3}& =\frac{\left( 2T-T^{2}\right) \left\vert \alpha
\right\vert ^{4}+2T\left\vert \alpha \right\vert ^{2}}{(\left\vert \alpha
\right\vert ^{2}+1)^{2}},  \tag{C3}
\end{align}%
which lead to%
\begin{equation}
\sum_{i}\delta ^{2}B_{i}=\frac{\left( 2+2T-T^{2}\right) \left\vert \alpha
\right\vert ^{4}+4\left\vert \alpha \right\vert ^{2}+2}{(\left\vert \alpha
\right\vert ^{2}+1)^{2}}.  \tag{C4}
\end{equation}

(3) Three variances for Charlie are$\allowbreak $%
\begin{align}
\delta ^{2}C_{1}& =\frac{\left\vert \alpha \right\vert ^{4}+\left(
1+T-\left( 1-T\right) \cos 2\phi \right) \left\vert \alpha \right\vert ^{2}+1%
}{(\left\vert \alpha \right\vert ^{2}+1)^{2}},  \notag \\
\allowbreak \allowbreak \delta ^{2}C_{2}& =\allowbreak \allowbreak \frac{%
\left\vert \alpha \right\vert ^{4}+\left( 1+T+\left( 1-T\right) \cos 2\phi
\right) \left\vert \alpha \right\vert ^{2}+1}{(\left\vert \alpha \right\vert
^{2}+1)^{2}},  \notag \\
\delta ^{2}C_{3}& =\frac{\left( 1-T^{2}\right) \left\vert \alpha \right\vert
^{4}+2\left( 1-T\right) \left\vert \alpha \right\vert ^{2}}{(\left\vert
\alpha \right\vert ^{2}+1)^{2}},  \tag{C5}
\end{align}%
which lead to

\begin{equation}
\sum_{i}\delta ^{2}C_{i}=\frac{\left( 3-T^{2}\right) \left\vert \alpha
\right\vert ^{4}+4\left\vert \alpha \right\vert ^{2}+2}{(\left\vert \alpha
\right\vert ^{2}+1)^{2}}.  \tag{C6}
\end{equation}

(4) Three covariances between Alice and Bob are%
\begin{align}
C\left( A_{1},B_{1}\right) & =-\frac{\sqrt{T}(\left\vert \alpha \right\vert
^{4}-\left\vert \alpha \right\vert ^{2}\cos 2\phi )}{(\left\vert \alpha
\right\vert ^{2}+1)^{2}},  \notag \\
C\left( A_{2},B_{2}\right) & =-\frac{\sqrt{T}(\left\vert \alpha \right\vert
^{4}+\left\vert \alpha \right\vert ^{2}\cos 2\phi )}{(\left\vert \alpha
\right\vert ^{2}+1)^{2}},  \notag \\
C\left( A_{3},B_{3}\right) & =-\frac{T\left\vert \alpha \right\vert ^{4}}{%
(\left\vert \alpha \right\vert ^{2}+1)^{2}}.  \tag{C7}
\end{align}

(5) Three covariances between Alice and Charlie are

\begin{align}
C\left( A_{1},C_{1}\right) & =\frac{\sqrt{1-T}(\left\vert \alpha \right\vert
^{4}-\left\vert \alpha \right\vert ^{2}\cos 2\phi )}{(\left\vert \alpha
\right\vert ^{2}+1)^{2}},  \notag \\
C\left( A_{2},C_{2}\right) & =\frac{\sqrt{1-T}(\left\vert \alpha \right\vert
^{4}+\left\vert \alpha \right\vert ^{2}\cos 2\phi )}{(\left\vert \alpha
\right\vert ^{2}+1)^{2}},  \notag \\
C\left( A_{3},C_{3}\right) & =-\frac{\left( 1-T\right) \left\vert \alpha
\right\vert ^{4}}{(\left\vert \alpha \right\vert ^{2}+1)^{2}}.  \tag{C8}
\end{align}

(6) Three covariances between Bob and Charlie are%
\begin{align}
C\left( B_{1},C_{1}\right) & =-\frac{\sqrt{T\left( 1-T\right) }(\left\vert
\alpha \right\vert ^{4}-\left\vert \alpha \right\vert ^{2}\cos 2\phi )}{%
(\left\vert \alpha \right\vert ^{2}+1)^{2}}\allowbreak ,  \notag \\
C\left( B_{2},C_{2}\right) & =-\frac{\sqrt{T\left( 1-T\right) }(\left\vert
\alpha \right\vert ^{4}+\left\vert \alpha \right\vert ^{2}\cos 2\phi )}{%
(\left\vert \alpha \right\vert ^{2}+1)^{2}},  \notag \\
C\left( B_{3},C_{3}\right) & =-\frac{T\left( 1-T\right) \left\vert \alpha
\right\vert ^{4}}{(\left\vert \alpha \right\vert ^{2}+1)^{2}}.  \tag{C9}
\end{align}

\end{document}